\documentclass[aps,twocolumn, superscriptaddress,groupedaddress,nofootinbib]{revtex4-1}
\usepackage[dvipsnames]{xcolor}
\usepackage[utf8]{inputenc}
\usepackage{url}
\usepackage{amsmath}
\usepackage{tikz}
\usepackage{amsfonts}
\usepackage{amssymb} 
\usepackage{bm}
\usepackage{graphicx}
\usepackage{array}
\usepackage{lipsum}
\usepackage{float}
\usepackage{multirow}
\usepackage{hhline}
\usepackage{tabularx}
\usepackage{subfigure}
\usepackage{graphicx}
\usepackage{afterpage}
\usepackage{longtable}
\usepackage{epstopdf}
\usepackage{MnSymbol}
\usepackage[utf8]{inputenc}

\usepackage{amsthm}

\newcommand{\Dc}{\Delta^\circ}

\definecolor{cornellRed}{HTML}{B31B1B}

\def\cV{\mathcal{V}}
\def\cR{\mathcal{R}}

\begin{document}
\title{On Axion Reheating in the String Landscape}
\author{James Halverson, Cody Long, Brent Nelson, and Gustavo Salinas}
\affiliation{Department of Physics, Northeastern University \\ Boston, MA 02115-5000 USA} 

\date{\today}

\begin{abstract}
We demonstrate that asymmetric reheating arises
 in a large ensemble of string compactifications
with many axions and gauged dark sectors. This phenomenon may help avoid
numerous cosmological problems that may arise if the sectors
were reheated democratically.
Distributions of couplings are presented for two classes
of axion reheatons, both of which exhibit very small couplings
to most of the gauge sectors. In one class, 
ratios of reheating couplings and also preferred gauge groups are frequently
determined by local regions in the string geometry.
\end{abstract}

\maketitle

\section{Introduction}

An inflationary epoch in the early universe provides solutions to the monopole, horizon, and flatness problems~\cite{PhysRevD.23.347, Linde:1981mu}. During the exponential expansion of
space caused by inflation, energy stored in fields other than the inflaton is quickly diluted, requiring a transfer of energy
to Standard Model particles at the end of inflation, known
as reheating \cite{ABBOTT198229,Dolgov:1982th,PhysRevLett.48.1437,Kofman:1994rk, Shtanov:1994ce}. More generally, scalar
fields other than the inflaton could be responsible for
reheating; henceforth, we refer to this field as the reheaton. There are many models of reheating, as well as non-perturbative generalizations such as preheating~\cite{Traschen:1990sw, Kofman:1994rk,Adshead:2015pva,GarciaBellido:1997wm,Cuissa:2018oiw}. Crucial to all are the couplings of the reheaton to the visible sector, as well as potential dark sectors. 
However, reheating
of dark sectors can lead to cosmological issues, such
as the overproduction of glueball dark matter 
\cite{Halverson:2016nfq, Halverson:2018olu}, which may often be exacerbated in
the presence of many gauged dark sectors. The problem
can be partially ameliorated  if the reheaton couples
more to some sectors than others; i.e., if reheating is asymmetric (see, e.g., \cite{Kolb:1985bf,Berezhiani:1995am,Adshead:2016xxj}). 

It has long been known that string theory gives rise to
gauged dark sectors (see, e.g., \cite{Gross:1984dd,Lerche:1986cx}), but in recent years their degree of relevance has strengthened significantly. For instance, the F-theory \cite{Vafa:1996xn, Morrison:1996pp}
geometry with $O(10^{272,000})$
flux vacua  \cite{Taylor:2015xtz} and the exact ensemble 
\cite{Halverson:2017ffz} of $\frac43 \times 2.96 \times 10^{755}$ F-theory geometries exhibit $33$ and
and $762 \pm 11$ gauged dark sectors, respectively. Together with \cite{Taylor:2017yqr}, these
comprise perhaps the largest known concrete regions of the landscape
of string vacua, and they motivate cosmologies with
large numbers of gauge sectors and axions.\footnote{When we refer to axions,
 we mean not only the pseudoscalars arising from string
 compactification that couple to the gauge sectors, but
 also more general axion-like particles.} 
Details of those cosmologies, and in particular whether they are realistic, depend crucially on reheating.

\vspace{.5cm}
We initiate a study of reheating in these constructions,
focusing on the case of axion reheating. 
We will calculate the couplings of axions to 
the various gauge sectors in the geometries of \cite{Taylor:2015xtz} and \cite{Halverson:2017ffz}; the former
is known as $B_\text{max}$, and the latter
is known as the Tree ensemble. We will demonstrate 
that axions that couple significantly to only a few gauge sectors
arise naturally in the ensemble, giving rise to asymmetric
reheating. We will show that such hierarchies often arise
from local patches in the string geometry. Specifically, in the case that the axion reheaton is defined to be
oriented along a gauge direction, we show that the sectors that are reheated to the highest temperatures are
that gauge direction, and also gauge sectors that intersect this sector; for fixed axion reheaton and Calabi-Yau, 
the other gauge sectors are reheated to a universal temperature. For axion reheatons that are random combinations of 
gauge-direction axions, we still find asymmetric reheating, but to a lesser degree. We also explain how this geometric result could be extended to a larger region of the string landscape.

While there has been some significant progress in understanding reheating in string theory (see e.g.~\cite{Kofman:2005yz,Cicoli:2010ha,Blumenhagen:2014gta}), there are important 
distinctions from the present work.
For instance, previous works have sometimes been limited to a few examples, and (more importantly) all in the case where the number of fields and gauge sectors is relatively small. In contrast, we study over a thousand geometries drawn from the Tree ensemble. The average number
of gauge sectors is $75$; this differs from the previously stated number of $762$ due to a truncation
made for computational reasons.
We also study reheating on $B_\text{max}$.
For a third regime of interest, the
reheaton could also arise from the open string sector. Asymmetric reheating may also arise there, likely due to
symmetries that charge open strings and therefore  bias some sectors over others. On the other hand, for an inflaton reheaton, these models are more likely to give rise to an $\eta$-problem that spoils slow-roll,
which the axion case avoids due  its shift symmetry.
We leave a study of open string reheatons in these ensembles to future work.

This paper is organized as follows. 
In Section \ref{sec:eft}
we present effective field theories of axion reheating and
their realization in F-theory. In Section \ref{sec:asymmetric}
we compute distributions of reheating ratios in
the Tree ensemble, and also in the F-theory
geometry with the most flux vacua.
In Section \ref{sec:gaugecorr} we study which
gauge groups are most likely to be reheated under
various assumptions. In Section \ref{sec:discuss}
we summarize our results and discuss further
aspects of the cosmology.

\clearpage

\section{Effective Field Theory \\ for Axion Reheating \label{sec:eft}}

We will consider an effective theory of $N$ axions $\phi^i$ and $P$ gauge sectors with field strengths $G_{\alpha \mu \nu}$ in F-theory, where the index $\alpha = 1, \dots, P$ runs over the various gauge groups. The general, canonically-normalized, two-derivative effective Lagrangian takes the form 
\begin{align}\label{eqn:cannorm}
& \mathcal{L} = -\frac{1}{2}\delta_{ij} (\partial^\mu \phi^i)(\partial_\mu \phi^j) - V(\phi)  \nonumber \\ 
& - \frac{1}{4}\sum\limits_\alpha  G_\alpha^{\mu\nu}G_{\alpha \mu \nu} - \sum\limits_\alpha c^{\alpha}_{\, i} \phi^i \tilde{G}_\alpha^{\mu\nu}G_{\alpha \mu \nu}\, ,
\end{align}
where $\tilde{G}_\alpha^{\mu \nu} = \epsilon^{\mu\nu\gamma\sigma} G_{\alpha \gamma \sigma}$ and $V(\phi)$ is the non-perturbative axion potential. We leave
all traces implicit.

Let us consider a candidate axion reheaton $\check{\phi}$. Our goal is to compute the couplings of $\check{\phi}$ 
to the various gauge sectors. In general there are many possibilities for $\check{\phi}$ and the one that is realized cosmologically is determined by inflationary dynamics. We will therefore address a simpler, but important question: 
if we can select a $\check{\phi}$ in order to attempt to reheat a single gauge sector $G_{\check{\alpha}}$, how much does $\check{\phi}$
necessarily couple to other gauge sectors? What is the
distribution of those couplings? In answering this question, we will uncover
a general geometric result which can be applied to more general cases. We consider a reheaton
$\check{\phi}$ directed along the $\check{\alpha}$-th gauge direction. 
We can in general write the axion-gauge boson interactions as
\begin{align}\label{eqn:aligned}
 -\mathcal{L}_{\text{int}} &= |g^{\check{\alpha}}| \hat{g}^{\check{\alpha}}_i \phi^i \tilde{G}_{\check{\alpha}}^{\mu\nu}G_{\check{\alpha} \mu \nu} + \sum\limits_{\alpha \neq \check{\alpha}} 
 |g^{\alpha}| \hat{g}^{\alpha}_{\, i} \phi^i \tilde{G}_\alpha^{\mu\nu}G_{\alpha \mu \nu}\nonumber \\
 & \equiv |g^{\check{\alpha}}| \check{\phi} \tilde{G}_{\check{\alpha}}^{\mu\nu}G_{\check{\alpha} \mu \nu} + \sum\limits_{\alpha \neq \check{\alpha}} 
 |g^{\alpha}| \hat{g}^{\alpha}_{\, i} \phi^i \tilde{G}_\alpha^{\mu\nu}G_{\alpha \mu \nu}\, ,
\end{align}
where $\hat{g}^\alpha$ are unit vectors. Our candidate reheaton is then $\check{\phi} = \hat{g}^{\check{\alpha}}_i \phi^i$. In this case we will say that
$\check{\phi}$ is oriented along $G_{\check{\alpha}}$.
Clearly $\check{\phi}$ couples to $G_{\check{\alpha}}$, but in general $\check{\phi}$ will appear in the second sum in Eq.~\ref{eqn:aligned} (which we might not expect, a priori), and will
therefore couple to other gauge groups, as well. Schematically, the effective interaction Lagrangian can be expanded as
\begin{align}\label{eqn:aligned2}
 -\mathcal{L}_{\text{int}} 
 & = c^{\check{\alpha}} \check{\phi} \tilde{G}_{\check{\alpha}}^{\mu\nu}G_{\check{\alpha} \mu \nu} + \sum\limits_{\alpha \neq \check{\alpha}} 
 (c^\alpha \check{\phi} + \dots )\tilde{G}_\alpha^{\mu\nu}G_{\alpha \mu \nu}\, ,
\end{align}
where we have only retained the $\check{\phi}$ axion-dependance in Eq.~\ref{eqn:aligned2}.
Our goal is then to compute the relative coupling strengths of $\check{\phi}$ to $G_{\check{\alpha}}$ and $G_\alpha$
with $\check{\alpha} \neq \alpha$, given by $c^{\check{\alpha}}$ and $c^\alpha$, respectively. From Eq.~\ref{eqn:aligned} we can see that $ c^{\check{\alpha}} = |g^{\check{\alpha}}|$. To read
off $c^\alpha$, the coupling of  $\check{\phi}$ to another other gauge group $G_\alpha$, we perform an $SO(N)$ basis transformation $\phi^i = M^{i}_{\, j} \varphi^j$, such that $\varphi^1 =\check{\phi}$, and then expand Eq.~\ref{eqn:aligned} in that basis. We then see that $c^\alpha =  |g^{\alpha}| \hat{g}^{\check{\alpha}} \cdot \hat{g}^{\alpha}$. 
We will focus on the ratio of the couplings $c^\alpha/c^{\check{\alpha}}$, and we therefore define the quantity of interest as
\begin{equation}\label{eqn:defR}
\cR_{\check{\alpha}}^\alpha = \frac{c^\alpha}{c^{\check{\alpha}}}= \frac{|g^{\alpha}| \hat{g}^{\check{\alpha}} \cdot \hat{g}^{\alpha}}{|g^{\check{\alpha}}|}
 = \frac{ g^{\check{\alpha}} \cdot g^{\alpha}}{|g^{\check{\alpha}}|^2}
\,.
\end{equation}
We will explore these couplings in F-theory, and uncover some interesting structure.

Let us now consider the data of such EFTs derived from F-theory. We will view these as IIb compactified on a K\"ahler threefold
$B$, with generalized 7-branes. After stabilizing complex structure moduli, the light
degrees of freedom of the $\mathcal{N} = 1$ effective field theory
are gravity, gauge sectors, and $h^{1,1}(B)$ K\"ahler moduli.
The latter are
written as:
\begin{equation}
T^i = \int\limits_{D_i} \left(\frac{1}{2} J \wedge J + i C_4\right)\equiv \tau^i + i\, \theta^i\, ,
\end{equation}
where the $D_i$ are a basis of divisors on $B$, $J$ is the K\"ahler form on $B$, and $C_4$ is the Ramond-Ramond four-form gauge potential.
The $\tau^i$ are volume moduli, which parametrize divisor volumes in $B$, while the $\theta^i$ are the associated axions.
With fixed volume moduli, the effective Lagrangian for
the axions and gauge fields contains the terms
\begin{align}\label{eqn:eff}
& \mathcal{L}  = -\frac{1}{2}K_{ij} (\partial^\mu \theta^i)(\partial_\mu \theta^j) - V(\theta)  \nonumber \\ 
& - \sum\limits_\alpha Q^{\alpha}_{\, i} \left( \tau^i F_\alpha^{\mu\nu}F_{\alpha \mu \nu} + 
 \theta^i \tilde{F}_\alpha^{\mu\nu}F_{\alpha \mu \nu}\right)\, .
\end{align}

The tree-level K\"ahler potential for the K\"ahler moduli is written as
\begin{equation}\label{eqn:kpot}
K = -2\, \text{log}(\cV)\, .
\end{equation}
Here $\cV$ is the volume of $B$ computed with the K\"ahler form $J = t_i \omega^i$, where $\omega^i$ is a basis for $H^{1,1}(B)$:
\begin{equation}\label{eqn:vol}
\cV = \frac{1}{6} \int_B J \wedge J \wedge J = \frac{1}{6}\kappa^{ijk}t_i t_j t_k\, .
\end{equation}
The $\kappa^{ijk} \equiv D_i \cdot D_j \cdot D_k$ are the triple intersection numbers of divisors. Eq.~\ref{eqn:kpot} is known to receive corrections in both
$\alpha^{\prime}$ and the string coupling $g_s$. As F-theory generically
contains 7-branes with $O(1)$ $g_s$ regions~\cite{Halverson:2016vwx}, such corrections are not well understood in F-theory, so we will take Eq.~\ref{eqn:kpot} as 
a model for the K\"ahler potential. 

The metric for the K\"ahler moduli then takes the form (for a derivation, see, e.g., \cite{Halverson:2017deq})
\begin{equation}\label{eqn:pot}
K_{ij} = \frac{1}{4}\left(-\frac{A_{ij}}{\cV} + \frac{t_i t_j}{2\cV^2}\right)\, ,
\end{equation}
where $A_{ij}$ is the inverse of $A^{ij} \equiv \kappa^{ijk}t_k = \text{vol}(D_i \cap D_j)$. 
The inverse K\"ahler metric will play
an important role in our analysis, and takes the form
\begin{equation}\label{eqn:invK}
K^{ij} = 4\left( -\cV A^{ij} + \tau^i \tau^j \right)\, .
\end{equation}

The couplings of the axions to the gauge bosons, given by the $Q^{\alpha}_{\, i}$ in Eq.~\ref{eqn:eff},
are determined by the wrappings of 7-branes on divisors: a 7-brane which carries a gauge sector and wraps a divisor $Q^{\alpha}_{i} D_i$ couples to the
linear combination of axions $Q^{\alpha}_{i} \theta^i$.  

Let us express $c^{\check{\alpha}}$, $c^\alpha$, and $\cR_{\check{\alpha}}^\alpha$ in terms of the geometric basis $\theta^i$. The kinetic term shown in Eq. (\ref{eqn:eff}) for the axion fields can be brought to the canonical form in Eq. (\ref{eqn:cannorm}) by diagonalizing the K\"ahler metric $K_{ij}$. Being symmetric and positive-definite, it can be decomposed as the matrix product $K= S^{\mathrm{T}} f^2 S = (fS)^{\mathrm{T}} (fS)$, with $S$ orthogonal and $f$ diagonal. Thus, the relation $\phi = (fS)\theta ~\Rightarrow~ \theta = (fS)^{-1} \phi$ follows. Plugging this into Eq.  (\ref{eqn:eff}) and canonically normalizing the gauge fields, we can see that the dot products in Eq.~\ref{eqn:defR} can be computed as
\begin{equation}
g^\alpha \cdot g^\beta= \tilde{Q}^{\alpha} \cdot K^{-1} \cdot \tilde{Q}^{\beta}\, ,
\end{equation}
where
\begin{equation}\label{eqn:normQ}
\tilde{Q}^\alpha = \frac{Q^\alpha}{4Q^{\alpha}_i \tau^i}\,.  
\end{equation}
The division by the volume factor in Eq.~\ref{eqn:normQ} is a result of the canonical normalization of the gauge fields. It is interesting to note that the $c^\alpha$ are invariant under
the K\"ahler parameter scaling $t_i \rightarrow \lambda t_i$, $\lambda \in \mathbb{R}$, as the $\tilde{Q}^{\alpha}$ scale with weight $-2$, due to the $\tau^i$ in the denominator of Eq.~\ref{eqn:normQ}, and $K^{-1}$ scales with weight $4$. This implies that the $c^\alpha$ are only functions of the angular coordinates in the K\"ahler cone, or the cone in which all curves in $B$ have positive volume.
It is convenient to define the variables $x^{\alpha\beta}$ as
\begin{equation}
x^{\alpha \beta} \equiv \frac{\text{vol}(D_\alpha \cap D_\beta)}{\text{vol}(D_\alpha)\times\text{vol}(D_\beta)}\, .
\end{equation}
We can then write
\begin{equation}\label{eq:selfcouple}
c^{\check{\alpha}} = \frac{1}{2}\sqrt{-\mathcal{V}x^{\check{\alpha} \check{\alpha}} + 1}\, ,
\end{equation}
and 
\begin{equation}\label{eq:othercouple}
c^{\alpha} = \frac{-\mathcal{V}x^{\check{\alpha} \alpha} + 1}{2\sqrt{-\mathcal{V}x^{\check{\alpha} \check{\alpha}} + 1}}\, ,
\end{equation}
and therefore
\begin{equation}\label{eqn:ratio}
\cR_{\check{\alpha}}^\alpha = \frac{x^{\check{\alpha} \alpha} - 1/\cV}{x^{\check{\alpha} \check{\alpha}} - 1/\cV}\, ,
\end{equation}
 Our goal will be to compute $\cR_{\check{\alpha}}^\alpha$ in an ensemble of F-theoretic examples.

Let us briefly comment on the potential for the axions. As mentioned above, the axions enjoy a continuous shift symmetry to all orders in perturbation theory, which is broken to a discrete shift symmetry by non-perturbative effects. As we will explain in the next section, the theories we will consider have large numbers of axions and condensing gauge sectors, whose low-energy dynamics will frequently result in gaugino condensates. The non-perturbative superpotential for the axions takes the schematic form
\begin{equation}\label{eqn:super}
W = W_0 + \sum\limits_{a = 1}^{P} A_a e^{-2 \pi Q^a_i (\tau^i _ i + i \theta^i)/C_2^a}\, ,
\end{equation}
where $W_0$ and the $A_a$'s are constants, and the $Q^a_i$ are the wrapping numbers of the 7-branes generating the gaugino condensates, with dual Coxeter numbers $C_2^a$. There can also be stringy ED3-instantons that do not correspond to gauge theory instantons (see, e.g., the review \cite{Blumenhagen:2009qh}), in which case $c_2^a \equiv 1$. The superpotential in Eq.~\ref{eqn:super}, the K\"ahler potential in Eq.~\ref{eqn:kpot}, and its derivatives enter the $\mathcal{N}=1$ supergravity potential to form the axion potential $V(\theta)$. Importantly, the axion masses are exponentially sensitive to the vacuum expectation values (vevs) of the K\"ahler moduli, and therefore to ensure that there exists an axion that can reheat the Standard Model to the required temperature, we need to ensure there is an axion whose corresponding saxion expectation value is not too large. We will later argue we always have at least one such axion. We will now review the construction and geometry of the largest-known explicit ensemble of F-theory geometries, known as the tree ensemble.

\subsection{The Tree ensemble}
The Tree ensemble is an ensemble of $4/3 \times 2.96 \times 10^{755}$ 
extra-dimensional geometries suitable for F-theory compactifications. They are toric bases that are generated by blowups of toric subvarieties on two particularly rich weak Fano toric threefolds, and form a connected network of geometries~\cite{Carifio:2017nyb}. Such blowups are expected to provide reasonable F-theory compactifications as they are finite distance in moduli space ~\cite{Hayakawa1995DEGENERATIONOC,Wang97onthe,Grassi1991} from well-understood geometries that have weak coupling limits. Each base is topologically distance, and while the ensemble itself is much too large for a brute force scan, a detailed study of the construction algorithm of such bases yields universal results in the physics. In particular, virtually all the bases have large gauge sectors and many axions. An analytic proof
shows that to high  probability $(\geq .999995)$, the minimal gauge algebra on each geometry is $G = E_8^{10} \times F_4^{18} \times U^9 \times F_4^{H_2} \times G_2^{H_3} \times A_1^{H_4}$, where $U \in \{G_2, F_4, E_6\}$ is a model-dependent gauge algebra, and the $H_i$ depend on the ray structure of the toric fan. An even stronger result
is obtained by taking $200$ random samples from
the ensemble: the average number of gauge factors is $762 \pm 11$,
and the average rank of the gauge group is $1609\pm17$.  The gauge sectors correspond to so-called non-Higgsable clusters~\cite{Morrison:2012np,Morrison:2012js,Grassi:2014zxa,Morrison:2014lca,Halverson:2015jua,Taylor:2015ppa,Halverson:2017ffz,Taylor:2017yqr}, whose gauge group cannot be Higgsed geometrically, i.e., via brane splitting. While additional gauge groups can be further tuned, we will mainly focus on non-Higgsable gauge groups, as they require no tuning whatsoever. However, one should note that a tuned three-family MSSM (albeit with an undetermined number of Higgs pairs) arises ubiquitously on the one of the weak Fano toric threefolds~\cite{Cvetic:2019gnh}.  An overwhelming fraction of these geometries contain inherently strongly-coupled 7-branes~\cite{Halverson:2017vde}. 
In the
absence of flux these sectors do not have matter, and thus will
confine, producing glueballs at low energies.

Let us briefly review the essential geometry of the Tree ensemble. Due to its toric nature the tree ensemble is combinatorial. A smooth weak Fano toric threefold, which is the starting point for our ensemble, is associated to a fine regular triangulation $\mathcal{T}$ of a 3d reflexive polytope $\Dc$. Such a $\mathcal{T}$ defines a fan $F$. Toric blowups correspond to adding ``exceptional'' rays to $F$ and subdividing its cones. The blowup procedure terminates when the exceptional rays reach a certain distance from the boundary of $\Dc$, as the sufficient condition to be at finite distance in moduli space is violated; see~\cite{Halverson:2017ffz} for further details,
and \cite{Halverson:2018xge} for a lengthy introduction to the
ensemble. Such a procedure has been informally coined as ``adding a tree'', as the exceptional rays look like a tree above $\Dc$. Toric points correspond to triangles in $\mathcal{T}$, and toric curves to edges, and both admit tree structures above them by adding rays above the corresponding edge or triangle. Each additional ray added provides an additional axion to the EFT, and adding higher and higher trees forces more and larger gauge groups.

\section{Asymmetric reheating \\ from local couplings}
\label{sec:asymmetric}

In order to compute a large number of examples in a reasonable amount of time, we restrict ourselves to geometries with $\lesssim 250$ axions. We will find that this is sufficient to recognize a general pattern in the coupling ratios, and we will demonstrate how the behavior scales with the number of axions. This provides evidence that our
results also hold in the bulk of the ensemble. To truncate the ensemble to a manageable number of axions, we only allow trees over of ten toric points. We consider $1260$ geometries drawn from the Tree ensemble (this particular number was determined by running the geometric analysis on our cluster for 24 hours). Each geometry consists of random trees added over ten random toric points. The number of axions ranges from 139 to 213, with the average number of axions being 188. The number of gauge groups range from 57 to 78, with an average number of 75. The rank of the gauge groups ranged from 194 to 352, with an average value of 315.

The couplings $c^\alpha$ are functions of the volume moduli, and their values therefore depend on details of moduli stabilization.  Moduli stabilization is intricate with even a few moduli~\cite{Cicoli:2008va}, and often requires balancing terms the perturbative K\"ahler potential against the non-perturbative superpotential in a delicate fashion. Given that these corrections are not well-understood in F-theory, and are only computed to low-order in string theory~\cite{Berg:2004ek, Berg:2005ja}, we will instead make the assumption that the volume moduli are stabilized in a region where certain non-perturbative effects may be 
safely neglected.
For instance, there are known non-perturbative $\alpha'$ corrections to the K\"ahler potential, such as worldsheet instantons that contribute schematically as
\begin{equation}
\Delta K \sim \sum\limits_n \frac{e^{-2\pi n \text{vol}(C)}}{\cV}\, ,
\end{equation} 
where $C$ is a two-cycle. Therefore, as a proxy for control of the non-perturbative corrections in $\alpha'$, we enforce that $\text{vol}(C) \geq 1$ for all curves $C$. This region was defined in~\cite{Demirtas:2018akl} as the stretched K\"ahler cone.\footnote{The stretched K\"ahler cone is not technically a cone, but instead a subregion within the K\"ahler cone at a fixed distance from the walls.}

\subsection{Sampling the K\"ahler cone}

In~\cite{Demirtas:2018akl} it was shown that toric fourfolds and their anticanonical hypersurfaces typically have very narrow K\"ahler cones, forcing some four-cycle volumes to be large and some axions nearly massless if one demands stabilization in the stretched
K\"ahler cone for reasons of control. We find the same structure for our toric threefolds. As a measure for the opening angle of the K\"ahler cone, we compute $\cos(\theta_{\text{min}})$, defined by
\begin{equation}\label{eq:minang}
	\cos (\theta_{\text{min}}) := \displaystyle\min_{a,b} \Biggl(\frac{M^a \cdot M^b}{|M^a||M^b|}\Biggr)\,.
\end{equation}
Here the $M^a$ are the generators of the Mori cone, dual to the K\"ahler cone. In our ensemble these values range from $-0.99$ to $-0.71$, with a mean of $-0.90$. This suggests that the angular space
of the Mori cone is quite wide, and therefore the angular space of the K\"ahler cone is quite narrow. We therefore do not expect much variation in the physics as we change the angle, while remaining in the stretched K\"ahler cone.

To demonstrate this, we select the geometry with the largest $\cos(\theta_\text{min})$ and sample the stretched K\"ahler cone via random walks. To begin such a walk, we use Mathematica's FindInstance method to find an initial point. While the inner workings of this method are not directly available to us, we find that this initial point has a large number of toric curves with $\mathcal{O}(1)$ area, and we therefore informally refer to this initial point as the apex of the stretched K\"ahler cone. We initiate the random walks from multiple starting points, beginning with the apex $v_0$, and then scaling outward to $\lambda v_0$, with $\lambda \in \{1,\dots , 10 \}$. We consider unit length step sizes, measured with respect to the toric curves areas. The direction of the step is drawn from a normal distribution centered at zero, with variance $1/N$. Allowing for $N_s = 1000$ steps  in each random walk, we restrict ourselves to remain within the stretched K\"ahler cone by requiring all curves to have volume greater than or equal to unity. As we are aiming to sample the angular space of the stretched K\"ahler cone, we scale each point to a constant volume slice at the end of the random walk. For each scaled point we compute the mean fractional difference (MFD) for the four cycles $\tau^i$ with respect to starting point of the random walk $\tau_i^0$, given by 
\begin{equation}
MFD(\tau) = \frac{1}{N_s}\left(\sum\limits_i \frac{|\tau_i - \tau_{i}^0|}{\tau_i + \tau_{i}^0} \right) \, .
\end{equation}

We find a maximum MFD of $\sim 0.1$, which implies the maximum average deviation of the four-cycle volumes is approximately $10\%$. Given the small amount of volume variation in this region, and the fact that the axion-gauge field couplings are invariant under $t_i \rightarrow \lambda t_i$, $\lambda \in \mathbb{R}$, for the purposes of reading off generic physics in the stretched K\"ahler cone we find it sufficient to examine the apex.

\subsection{Results of the scan}

\begin{figure}
\includegraphics[width=.5\textwidth]{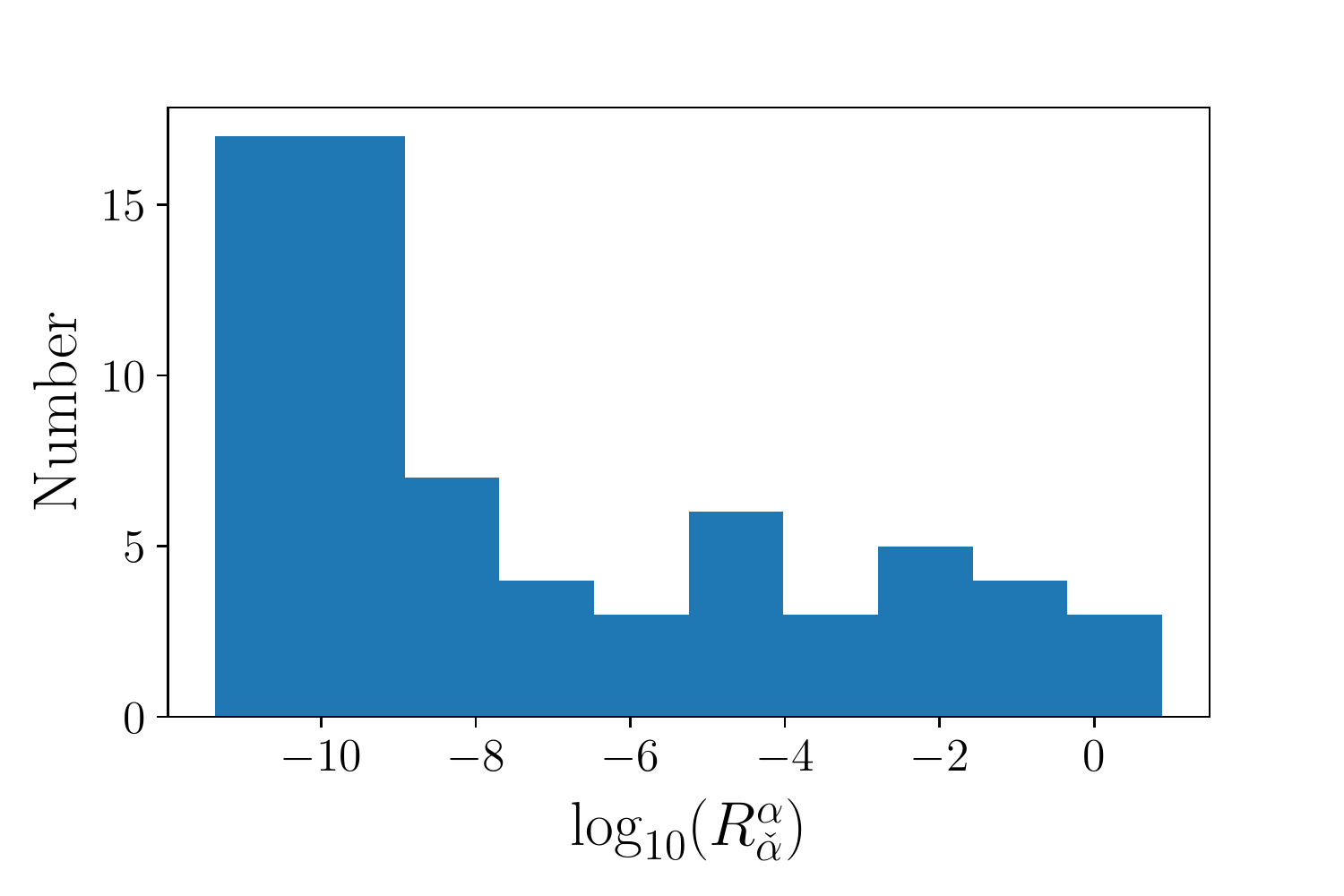}
\caption{
$\cR_{\check{\alpha}}^\alpha$ in a single geometry, for a single choice of $\check{\phi}$. In this particular case $\check{\phi}$ is aligned along an $E_8$ gauge group. We see that while there are a few additional sectors that couple to $\check{\phi}$ with similarly large coupling, most sectors have a much smaller coupling to $\check{\phi}$.
}
\label{fig:single}
\end{figure}

We compute the ratios $\cR_{\check{\alpha}}^\alpha$ in our ensemble of tree geometries. Let us first consider a single choice of $\check{\phi}$ in a single geometry. We show the distribution for $\cR_{\check{\alpha}}^\alpha$ in this example in Fig.~\ref{fig:single}. In this example there are a few sectors $G_\alpha$ that couple to the reheaton on the same order as the coupling to $G_{\check{\alpha}}$, while the rest of the couplings are significantly smaller by many orders of magnitude. This observation is in fact general throughout our ensemble.

\begin{figure}[t]
\includegraphics[width=.5\textwidth]{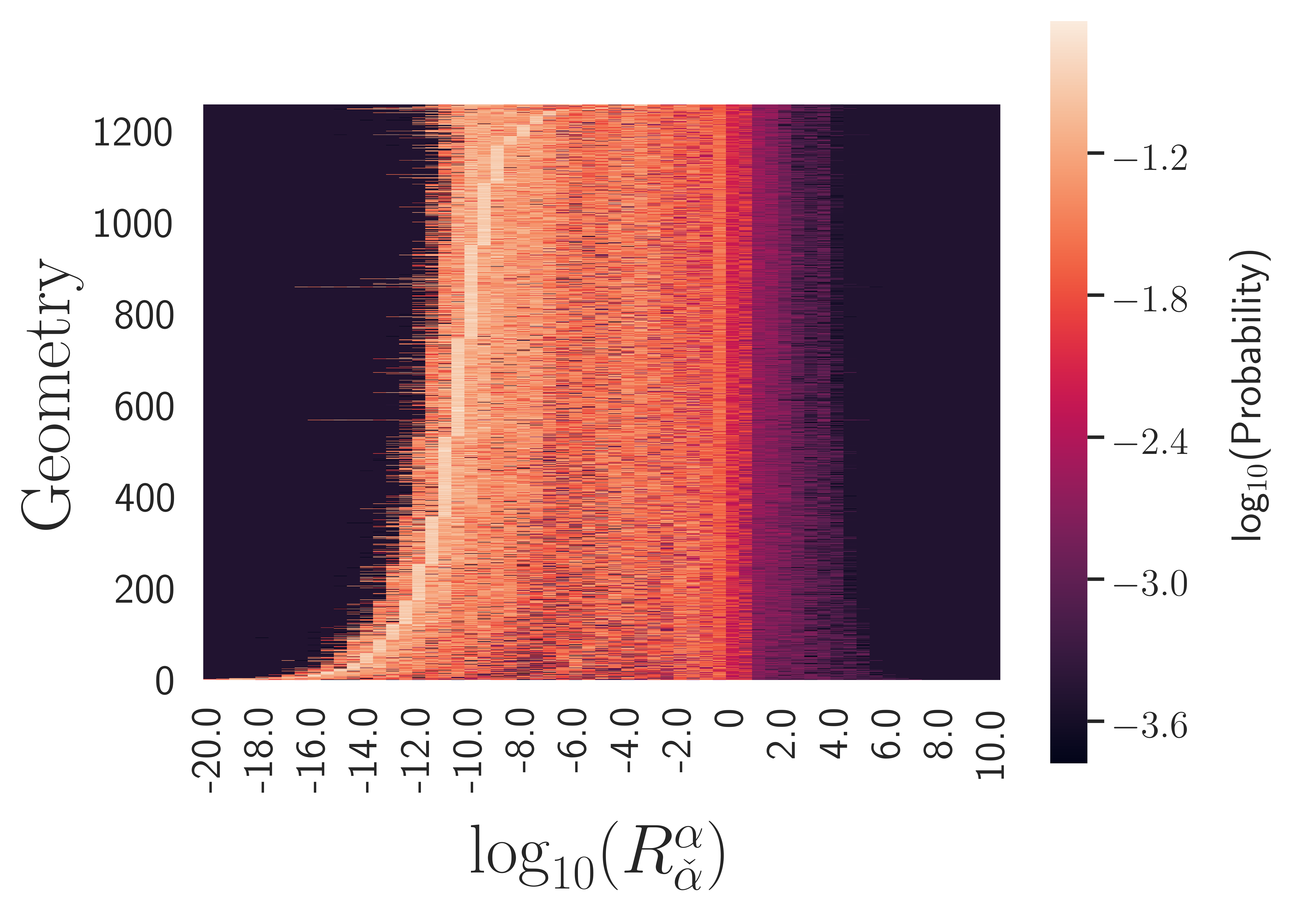}
\caption{
The distribution of $\cR_{\check{\alpha}}^\alpha$ from our scan over $1260$ F-theory geometries. Each horizontal line corresponds to a single geometry, and the data in a single horizontal line is the flattened distribution of all $\cR_{\check{\alpha}}^\alpha$ for that geometry. 
}
\label{fig:geom}
\end{figure}

We plot the bulk results from all the geometries in our scan in a heat map, shown in Fig.~\ref{fig:geom}. Specifically, each horizontal line in the heat map is a single geometry, and for each choice of gauge-oriented reheaton $\check{\phi}$ we compute $\text{log}_{10}\left(\cR_{\check{\alpha}}^\alpha\right)$
for all other gauge sectors $\alpha$.  The heat distribution is then probability to find a given $\text{log}_{10}\left(\cR_{\check{\alpha}}^\alpha\right)$. 

The first notable feature of these distributions is that the peak of a typical distribution is
around $\sim10^{-11}$, which suggests that, given a reheaton $\check{\phi}$ oriented along a gauge group $G_{\check{\alpha}}$, the relative strength of the couplings of $\check{\phi}$ to
other gauge sectors is usually negligible, down by over ten
orders of magnitude.  However, there is a tail towards larger values. From the right tail we can see that the fraction of $\mathcal{O}(1)$ $\cR_{\check{\alpha}}^\alpha$ is $\sim 0.01$, and that these ratios extend to even larger values, reaching into $\mathcal{O}(10^{6})$. In these cases, even though one attempts to arrange for a reheaton $\check{\phi}$ to reheat a particular sector, $\check{\phi}$ necessarily couples to other sectors with much stronger coupling. The conclusion is that for a $\check \phi$ chosen
to be oriented in a gauge direction, it will couple very weakly to most
of the gauge sectors in the compactification (recall there are $75$ on average), but will couple non-trivially to a few.\footnote{This result will soon be discussed in light of
common expectations about democratic reheating due to
global closed string effects.}
This feature is what we call asymmetric reheating.

\begin{figure*}
\includegraphics[origin=c,width=.47\textwidth]{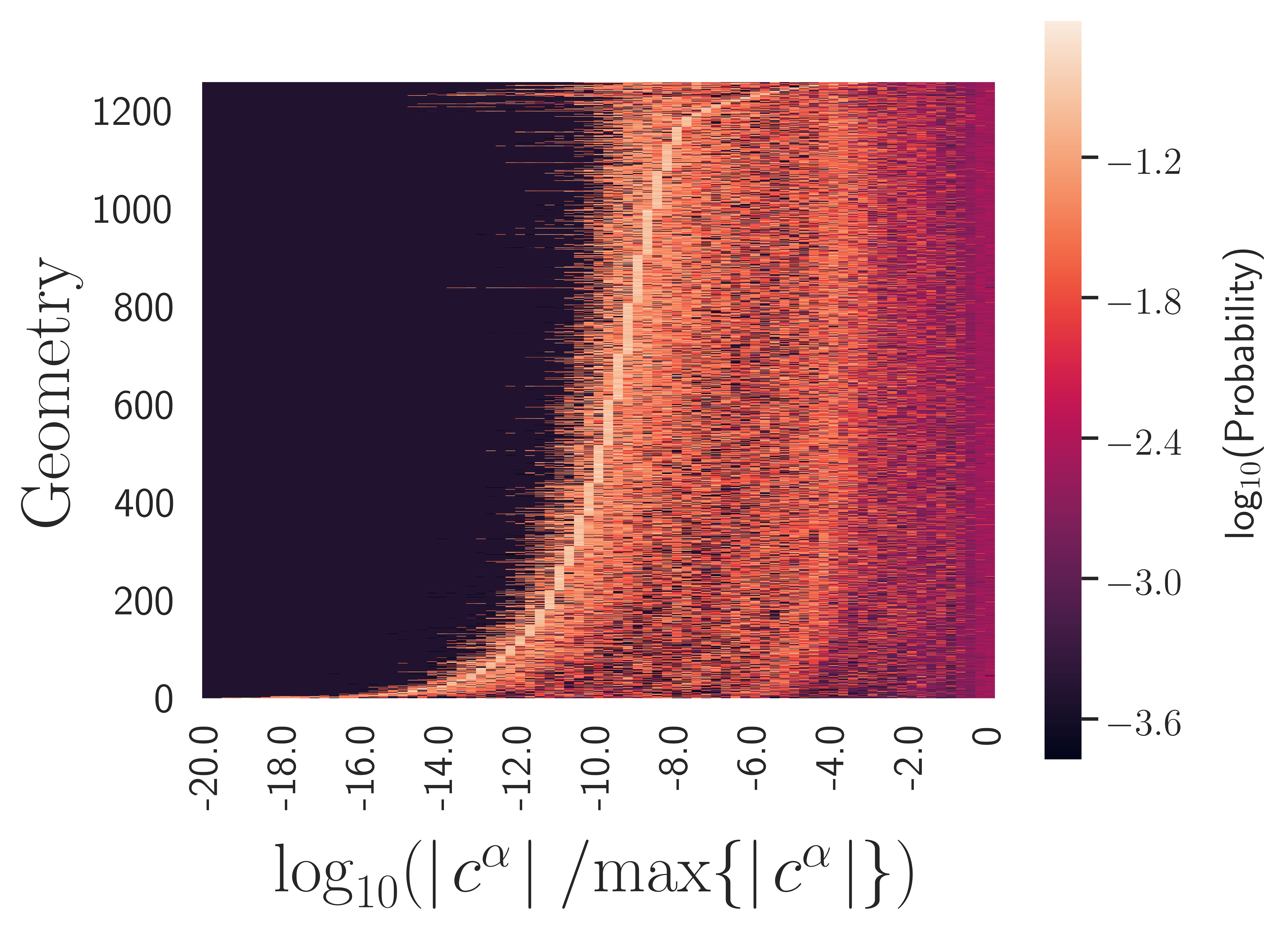}
\hspace{0.5cm}
\includegraphics[origin=c,width=.47\textwidth]{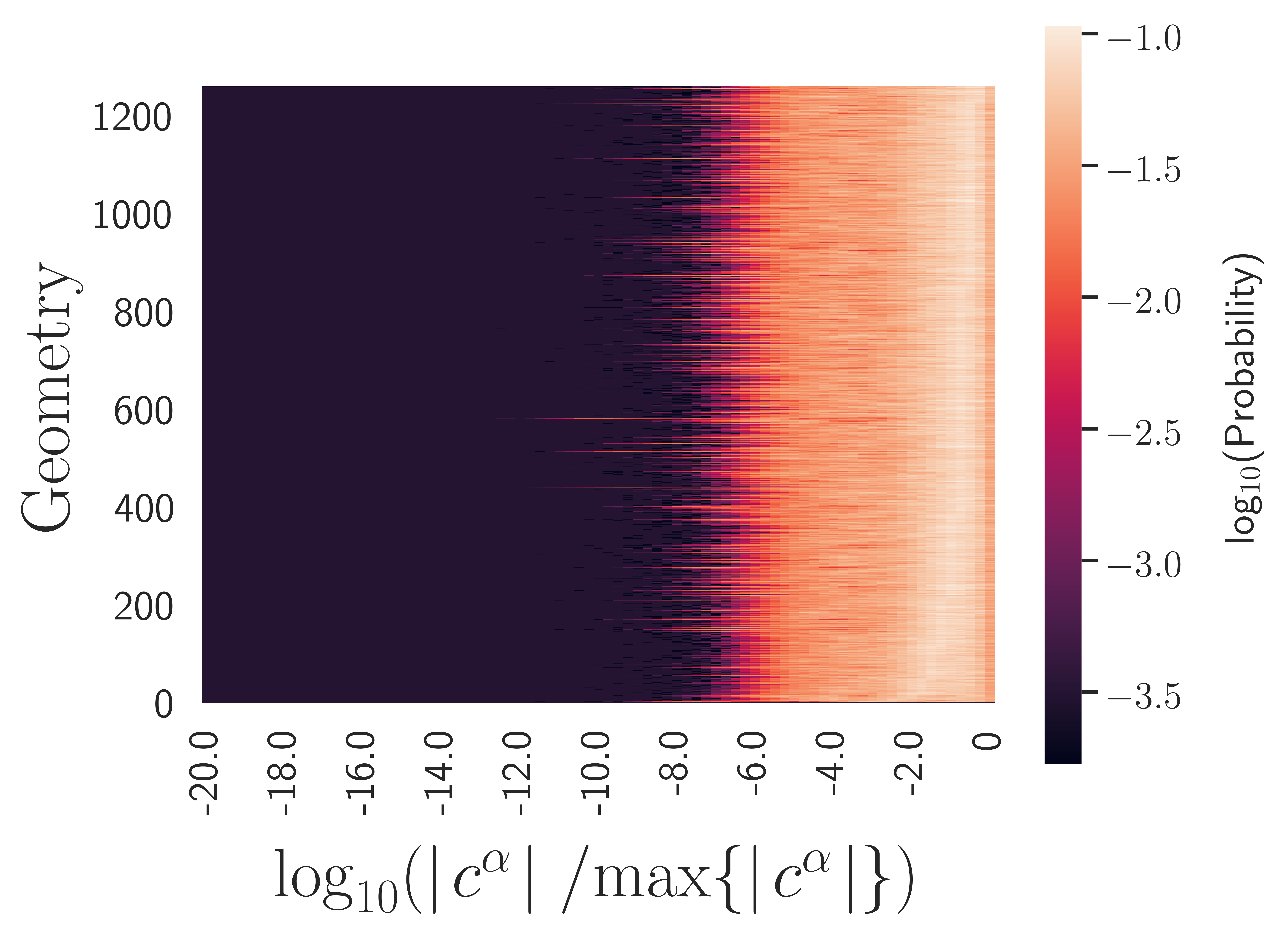}
\caption{\emph{Left:}
The bulk distribution of $\text{log}_{10}\left(|c^\alpha|/\text{max}\{|c^\alpha|\}\right)$ for the $\check{\phi}$ oriented along a single gauge sector. 
\emph{Right:} The bulk distribution of $\text{log}_{10}\left(|c^\alpha|/\text{max}\{|c^\alpha|\}\right)$ for random choices of $\check{\phi}$. This data
suggests that the oriented reheatons $\check{\phi}$ have much more asymmetric couplings that a random choice of $\check{\phi}$.}

\label{fig:geomvsrando}
\end{figure*}

The above data captures the necessary minimal couplings of an axion reheaton, oriented along a single gauge direction, to other gauge sectors. However, cosmological dynamics
could select other reheaton directions, and we should therefore contrast to another limiting point: random reheaton directions.  In order to compare to more general reheaton directions, we consider the quantity $|c^\alpha|/\text{max}\{|c^\alpha|\}$, as  $\cR_{\check{\alpha}}^\alpha$ is not well-defined if $\check{\alpha}$ does not correspond to a single gauge direction. To generate such random directions, we draw the entries of $\check{\phi}$ from a normal distribution centered around zero, with variance $1/h^{1,1}=1/N$. In the right hand side of Fig.~\ref{fig:geomvsrando} we plot the base-10 logarithm of $|c^\alpha|/\text{max}\{|c^\alpha|\}$ for random axion-reheaton directions. Each horizontal line in the plot is a single geometry, and the heat distribution is the base-10 logarithm of the probability to find a given $\text{log}_{10}\left(|c^\alpha|/\text{max}\{|c^\alpha|\}\right)$. In the left hand side of the same figure we also show $|c^\alpha|/\text{max}\{|c^\alpha|\}$ from the gauge directions to contrast with the random directions.

 We find that the couplings from the random directions are much less asymmetric that the oriented ones:
the peak of the distribution for random reheatons on the right-hand side of
Figure \ref{fig:geomvsrando} is close to $c/c_{max}\sim 1$
and has probability $\sim 1/10$,
indicating $O(1)$ couplings to about $10\%$ of the gauge
sectors. This should be contrasted to the left-hand side,
the distribution for gauge-direction reheatons, where we see
that the probability of an $O(1)$ coupling is $\sim 1/100$,
indicating $O(1)$ couplings to about $1\%$ of the gauge
sectors. This result for random reheatons is consistent
with expectations from effective field theory.

Another useful comparison arises from considering a random effective field theory of $N$-axions, with a Lagrangian of the form given in Eq. \ref{eqn:eff}. We draw the entries of $K_{ij}$ from a Wishart distribution, constructed from a normal distribution centered around zero, with variance $1/N$. In addition, the matrix $Q$ is taken as random with entries drawn from (i) a normal distribution centered around zero, with variance $1/N$, (ii) random integers or (iii) sparse random integers. In all three cases, the result is a distribution similar to the one shown in Fig.~\ref{fig:geomvsrando} (right), but with significantly larger entries: the peaks of the distributions are closer to unity. This indicates that a generic random EFT clearly differs from the random reheaton direction studied above.

\vspace{1cm}
In summary, we conclude that some degree of asymmetric
reheating is generic in this string ensemble. The
extent to which the reheating is asymmetric is determined by how much a reheaton is aligned with a gauge eigenstate direction vs. a random combination of those directions. In the former
case, we see $O(1)$ couplings to $\sim 1\%$ of the gauge
sectors, and in the latter case we see $O(1)$ couplings
to $\sim 10 \%$ of the gauge sectors.  Given a large
number (e.g. $O(75)$, as here) of gauge sectors, one expects that obtaining a realistic
cosmology requires some degree of asymmetric reheating. 
Our results demonstrate that models of axion reheating can potentially give rise to realistic cosmologies via
asymmetric reheating, particularly in the case that the reheaton is aligned along a single gauge group.

\subsection{Geometric origin of asymmetric reheating}

Let us now discuss the origin of this hierarchy of couplings
that lead to asymmetric reheating. 
Recall the definition of $\cR_{\check{\alpha}}^\alpha$, given in Eq.~\ref{eqn:ratio}. The numerator and denominator each have a local piece that depends on the intersection of $D_\beta$ ($\beta \in \{\check{\alpha}, \alpha \}$) and $D_{\check{\alpha}}$, given by their respective $x^{\alpha \beta}$, as well as a non-local additive factor of $1/\cV$. A priori it is not clear which piece, if either, should dominate; for instance both have the same volume scaling under $t \rightarrow \lambda t$. However, we find that the approximation
\begin{equation}\label{eqn:approx}
 \cR_{\check{\alpha}}^\alpha \approx \frac{x^{\check{\alpha} \alpha}}{x^{\check{\alpha} \check{\alpha}}}\, , 
\end{equation}
is highly accurate first-order approximation; that is, when $\cR \gtrsim 0.01$,  only the local pieces matter when $D_{\check{\alpha}}$ and $D_{\alpha}$ intersect, and the couplings are highly suppressed when they do not. 

We arrived at this conclusion by an analysis of 
$\cR_{\check{\alpha}}^\alpha$ using a neural network\footnote{Neural
networks are excellent function approximators, and are central
to recent efforts to study the string landscape using
machine learning; see, e.g., \cite{He:2017aed, Ruehle:2017mzq, Carifio:2017bov}.} using a strategy that we will
call input-dropout. The goal of this analysis is to understand
which variables are critical in predicting 
$\cR_{\check{\alpha}}^\alpha$, as well as to potentially discover correlations between such variables. Our network is a fully
connected feed-forward neural network implemented in
PyTorch, with five layers, each with width 100.
We initially trained the 
network to predict $\cR_{\check{\alpha}}^\alpha$ given
several input variables, which were the gauge group reheaton, the additional gauge groups, the graph distance between the various divisors (defined with respect to the toric fan), the overall volume of $B$, the volumes of all toric divisors, as well as the volume combinations $x^{\alpha \beta}$ (which is enough to implicitly determine the areas of the toric curves as well). These
additional input variables were included since they might
provide an alternative understanding of $\cR_{\check{\alpha}}^\alpha$. 

After
training for 10 epochs, the mean squared error
loss was $0.013$, indicating very
accurate predictions. However, often only a subset of
the input variables are important for accurately predicting
the output, and this can be tested in a simple way by
input-dropout: systematically removing some of the inputs
and checking whether the neural network still makes accurate
predictions. Removing all of the variables except 
$x^{\alpha \beta}$, we see that the network still predicts
$\cR_{\check{\alpha}}^\alpha$ with a mean squared error
loss of $0.09$
after training for 10 epochs. By contrast, if
those input variables are removed and all others are left
intact, the network only predicts with mean squared error
loss of $0.78$
after the same number of epochs. This large jump in mean square error loss suggests that any additional interesting or surprising correlations that could determine $\cR_{\check{\alpha}}^\alpha$ are intricate, and if present may require machinery beyond a simple neural net to uncover.

\begin{figure}[t]
\includegraphics[width=.5\textwidth]{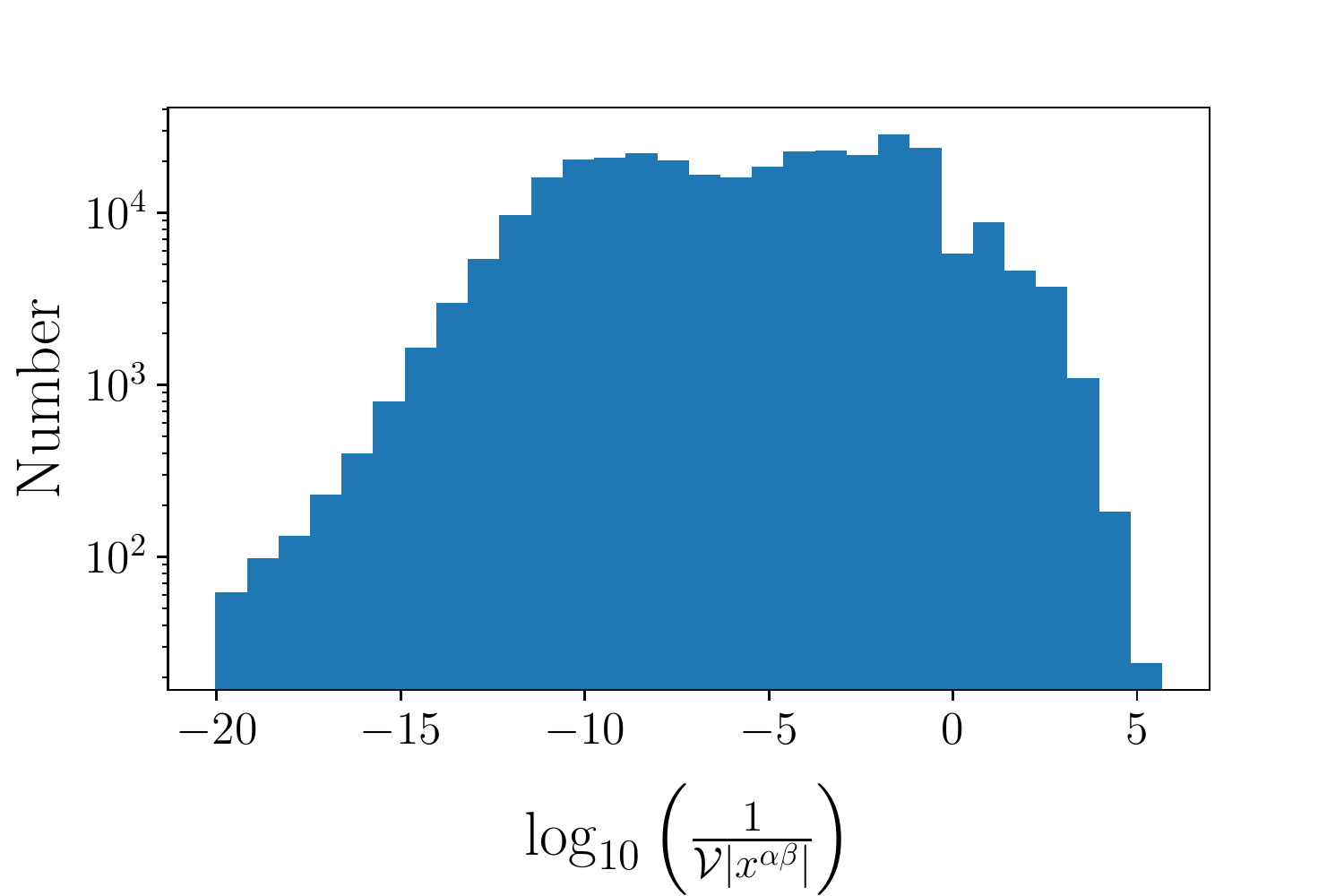}
\caption{
The ratios of the magnitudes of the inverse overall volume $1/\cV$ to the $|x^{\alpha \beta}|$, where $\alpha$ and $\beta$ range over all divisors with non-trivial intersection. This number is usually small, providing strong evidence that the ``non-local'' $1/\cV$ contribution to the axion-gauge couplings is negligible.
}
\label{fig:volvscurves}
\end{figure}

We therefore have the highly accurate approximation 
\begin{equation}\label{eqn:approx2}
 \cR_{\check{\alpha}}^\alpha \approx \frac{\text{vol}(D_{\check{\alpha}} \cap D_\alpha)}{\text{vol}(D_{\check{\alpha}} \cap D_{\check{\alpha}})} \times \frac{\text{vol}(D_{\check{\alpha}})}{\text{vol}(D_\alpha)}\, .
 \end{equation} 
 In retrospect, this result could have been derived instead
 by simply looking at the distribution of $1/(\cV |x^{\alpha \beta}|)$
 presented in Fig.~\ref{fig:volvscurves}, from which one concludes that
 the $1/\cV$ contribution is often negligible compared to that of $x^{\alpha \beta}$.
However, from Fig.~\ref{fig:volvscurves} we see that this approximation breaks down
 in about $15\%$ of the cases. Let us analyze the behavior in the regime of greatest interest, when the $\cR_{\check{\alpha}}^\alpha$ is significant ($\gtrsim 0.01$). Here we find 
 that $1/(\cV |x^{\check{\alpha} \alpha}|) < 0.1$ in $\{87\%, 89\%, 95\% \}$ of the cases when $\cR_{\check{\alpha}}^\alpha > \{10^{-2}, 10^{-1}, 1\}$, respectively, allowing us to approximate the numerator by $x^{\check \alpha \alpha}$. Expanding the denominator to
 leading order in $\frac{1}{\cV x^{\check{\alpha} \check{\alpha}}}$, we have
 \begin{equation}\label{eqn:compare}
 \cR_{\check{\alpha}}^\alpha \approx  \frac{x^{\check{\alpha} \alpha}}{x^{\check{\alpha} \check{\alpha}} - 1/\cV} \simeq  \frac{x^{\check{\alpha} \alpha}}{x^{\check{\alpha} \check{\alpha}}}\left(1 +\frac{1}{\cV x^{\check{\alpha} \check{\alpha}}} \right),
 \end{equation}
which corrects the local picture \eqref{eqn:approx}.
 To know how robust the local picture is, we would like to compute the size of the correcting factor. We find that $\frac{1}{\cV x^{\check{\alpha} \check{\alpha}}} < 0.1$ in $82\%$ of the cases with  $\cR_{\check{\alpha}}^\alpha \gtrsim 10^{-2}$, which leads to a correction factor $\leq 1.1$. 
 That is, in $82\%$ of these mentioned cases, the approximation \eqref{eqn:approx} is valid to within $10\%$.
 
\begin{figure}[t]
\begin{tikzpicture}[scale=1]
\draw[thick,color=Black] (0,0) -- (3,0) -- (0,3) -- cycle;
\draw[thick,color=Black] (0,.5) -- (2.5,.5);
\draw[thick,color=Black] (0,1) -- (2,1);
\draw[thick,color=Black] (0,1.5) -- (1.5,1.5);
\draw[thick,color=Black] (0,2) -- (1,2);
\draw[thick,color=Black] (0,2.5) -- (.5,2.5);
\draw[thick,color=Black] (.5,0) -- (.5,2.5);
\draw[thick,color=Black] (1,0) -- (1,2);
\draw[thick,color=Black] (1.5,0) -- (1.5,1.5);
\draw[thick,color=Black] (2,0) -- (2,1);
\draw[thick,color=Black] (2.5,0) -- (2.5,.5);
\draw[thick,color=Black] (0,2) -- (.5,2.5);
\draw[thick,color=Black] (0,1) -- (1,2);
\draw[thick,color=Black] (0,0) -- (1.5,1.5);
\draw[thick,color=Black] (1,0) -- (2,1);
\draw[thick,color=Black] (2,0) -- (2.5,.5);
\draw[thick,color=Black] (0,1.5) -- (.5,2);
\draw[thick,color=Black] (0,.5) -- (1,1.5);
\draw[thick,color=Black] (.5,0) -- (1.5,1);
\draw[thick,color=Black] (1.5,0) -- (2,.5);
\fill (0,0) circle (.5mm); \fill (0,.5) circle (.5mm); \fill (0,1) circle (.5mm);
\fill (0,1.5) circle (.5mm); \fill (0,2) circle (.5mm); \fill (0,2.5) circle (.5mm);
\fill (0,3) circle (.5mm);
\fill (.5,0) circle (.5mm); \fill (.5,.5) circle (.5mm); \fill (.5,1) circle (.5mm);
\fill (.5,1.5) circle (.5mm); \fill (.5,2) circle (.5mm); \fill (.5,2.5) circle (.5mm);
\fill (1,0) circle (.5mm); \fill (1,.5) circle (.5mm); \fill (1,1) circle (.5mm);
\fill (1,1.5) circle (.5mm); \fill (1,2) circle (.5mm); 
\fill (1.5,0) circle (.5mm); \fill (1.5,.5) circle (.5mm); \fill (1.5,1) circle (.5mm);
\fill (1.5,1.5) circle (.5mm); 
\fill (2,0) circle (.5mm); \fill (2,.5) circle (.5mm); \fill (2,1) circle (.5mm);
\fill (2.5,0) circle (.5mm); \fill (2.5,.5) circle (.5mm);
\fill (3,0) circle (.5mm);
\draw[blue,fill=blue] (1,1) circle (.7ex);
\draw[red,fill=red] (1,1.5) circle (.7ex);
\draw[red,fill=red] (1.5,1.5) circle (.7ex);
\draw[red,fill=red] (1.5,1) circle (.7ex);
\draw[red,fill=red] (1,.5) circle (.7ex);
\draw[red,fill=red] (.5,1) circle (.7ex);
\draw[red,fill=red] (.5,.5) circle (.7ex);

\end{tikzpicture}

\caption{An example of the graphical nature of determining gauge sectors that couple significantly to a gauge-oriented axion reheaton. The blue dot in the middle corresponds to
the divisor $D_{\check{\alpha}}$ supporting the gauge group the reheaton is oriented along, and the red dots correspond to the divisors $D_{\alpha}$ that support gauge sectors that can have significant coupling to the reheaton, determined by the fact that $D_{\check{\alpha}} \cap D_{\alpha} \nsim 0$, as they share an edge in the toric fan. }
\label{fig:bigfacetbigone1}
\end{figure}
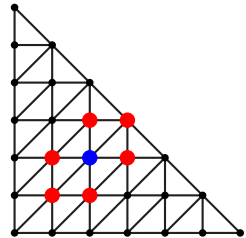

\vspace{.5cm}
The general lesson that we have learned in this analysis is that the local intersection data determines the axion-gauge couplings to high accuracy,
for axions oriented along gauge directions. In particular, if we consider a reheaton $\check{\phi}$ oriented along a gauge field supported on a divisor $D_{\check{\alpha}}$, then we only expect $\check{\phi}$ to couple significantly to gauge groups supported on divisors $D_{\alpha}$ with $D_{\check{\alpha}} \cap D_{\alpha} \nsim 0$. This fact reduces the problem of determining the potential sectors that $\check{\phi}$ can significantly couple to to a simple graph problem: given a $D_{\check{\alpha}}$, determine the divisors $D_{\alpha}$ such that $D_{\check{\alpha}} \cap D_{\alpha} \nsim 0$. Of course, this data is completely specified by whether the points corresponding to $D_{\check{\alpha}}$ and $D_{\alpha}$ share an edge in the toric fan. A simple example of this is shown is Fig~\ref{fig:bigfacetbigone1}. The blue dot corresponds to $D_{\check{\alpha}}$, while the red dots correspond to the $D_{\alpha}$ that intersect $D_{\check{\alpha}}$ non-trivially. 

This geometric observation opens the door for another possibility for engineering axion reheating into a gauge group $G$ that is supported on a divisor $D$. By arranging for an axion-reheaton $\check{\phi}$ that is oriented along a divisor $D_R$ that does not carry a gauge group, but instead intersect $D$ non-trivially. One then expects $\check{\phi}$ to couple fo $G$ non-trivially, due to the correspondence between intersections and significant couplings. If one could arrange for
a $D_R$ that does not carry a gauge group and that  intersects only a few gauge groups divisors, it would provide an additional candidate scenario for asymmetric axion reheating.

\subsection{Scaling with the number of axions}
\begin{figure}[t]
\includegraphics[width=.5\textwidth]{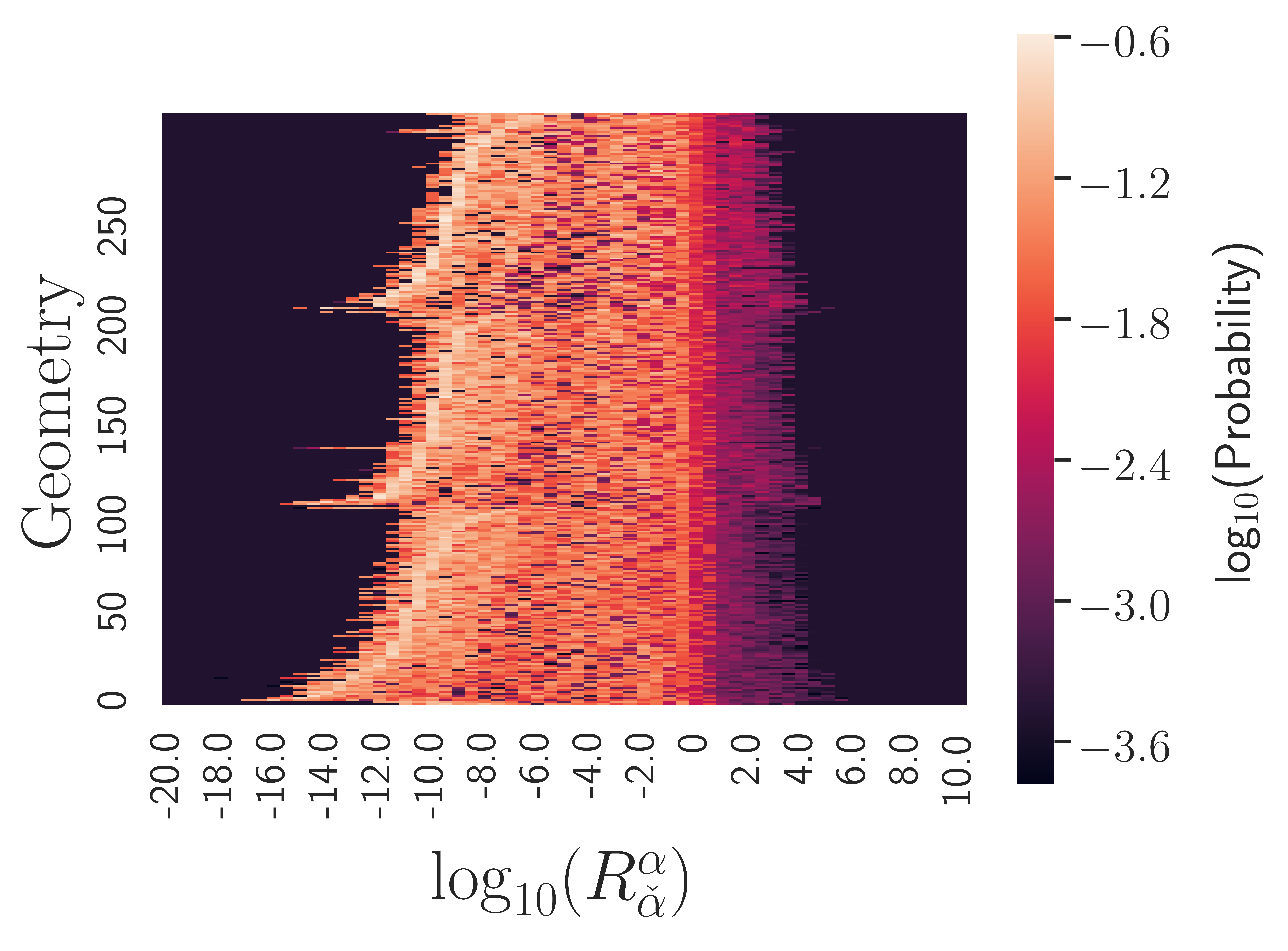}
\caption{
The distribution of $\cR_{\check{\alpha}}^\alpha$ from for 100 geometries with a) three trees (bottom third), b) six trees (middle third), and c) ten trees (top third). This plot provides strong evidence for universality of asymmetric axion-gauge sector couplings, independent of $N$. For three trees, the gauge group rank ranges from 70 to 247, and the number of axions ranges from 61 to 91. For six trees, the gauge group rank ranges 117 to 332, and the number of axions ranges from 102 to 142.
}
\label{fig:Nscaling}
\end{figure}

Our analysis thus far has concentrated on an ensemble generated by placing random trees over ten points, on a weak Fano toric variety, yielding compactification geometries with $\mathcal{O}(200)$ axions. However, a generic geometry in the Tree ensemble will have 72 trees over points, and $\mathcal{O}(1000)$ axions. It is therefore prudent that we analyze the scaling behavior of our result with the number of axions and gauge groups. To do so, we fix the number of trees at three and six, and for each draw 100 random geometries from the Tree ensemble, and perform the same analysis. For three trees, the gauge group rank ranges from 70 to 247, and the number of axions ranges from 61 to 91. For six trees, the gauge group rank ranges 117 to 332, and the number of axions ranges from 102 to 142. The results are shown in Fig~\ref{fig:Nscaling}, where we have plotted the distribution of $\cR_{\check{\alpha}}^\alpha$ for 100 geometries with three trees (bottom third of the plot), 100 with six (middle third), and 100 with ten (top third), drawn randomly from our previous analysis. The results demonstrate universality of $\cR_{\check{\alpha}}^\alpha$, regardless of the number of axions. This provides strong evidence that our results should hold at large $N$. 

This universality is a quite interesting feature, and suggests that much of the observed structure is likely coming from the underlying polytope, instead of the trees placed upon it. It would be interesting to better understand its origin.

\subsection{Bounds from nucleosynthesis}

We have studied in detail distributions of
the ratio of couplings of an axion
to two different gauge sectors, which are
central to perturbative asymmetric reheating.

However, we have not yet considered
the constraint on the reheat
temperature of the visible sector due
to nucleosynthesis. While a detailed study
of this important issue is beyond the scope
of this work, our goal is to demonstrate
that there is at least one axion that reheats
a gauge sector to above the required
temperature. Assuming instantaneous
decay of the axion and subsequent
thermalization,
\begin{equation}
T_{\text{rh}} = \left[\left(\frac{8}{90}\pi^3 g_\star\right)^{-\frac12} \, M_P\, \Gamma_a \right]^\frac12,
\end{equation}
where $g_\star$ is the number of active
degrees of freedom in the Standard Model visible sector. Since all of our reheating
couplings satisfy $c_{\check{\alpha}} > 1/M_P$,
we have $\Gamma_a \simeq c_\alpha^2 m_a^3 \gtrsim m_a^3/M_P^2$. Famously, due to the
cosmological moduli problem, rates of order
$m_a^3/M_P^2$ lead to $T_\text{rh} > T_\text{BBN} \simeq 5 \, \text{MeV}$ when $m_a \gtrsim 50\, \text{TeV}$.

We may check whether
this bound is satisfied via
the leading order
(in saxion VEV)
axion mass from equation (4.9) of \cite{Demirtas:2018akl},
\begin{equation}
m_a = \left(\frac{32 \pi^3\, \tau W_0}{\mathcal{V}^2 f^2}\, e^{-2\pi\tau/C_2}\right)^\frac12,
\end{equation}
where $f$ is
the axion decay constant that appears in
the $\text{cos}(2\pi a/f)$ in the potential
and $\tau$
is the four-cycle volume associated with the axion. To estimate
the typical $f$ we compute
the diagonal entries of the K\"ahler metric $K_{ii}$ in our ensemble.
Taking $f=10^{-7}$, $W_0=1$, $C_2=2$ for
$SU(2)$, and $\mathcal{V}=10^{11}$ (as is typical
in our studies), $m_a>50\text{TeV}$ is
satisfied for $\tau < 17$; for $E_8$,
which has $C_2=30$, the constraint is
is satisfied for $\tau < 270$. The latter
is quite easy to satisfy in the Tree ensemble,
but even in the former case we have at
least one four-cycle with $\tau < 17$
for each geometry.

We are therefore confident that a sufficient
reheat temperature can be achieved in
our ensemble.

\subsection{F-theory geometry with the most flux vacua}
\begin{figure}[t]
\includegraphics[angle=0,width=.5\textwidth]{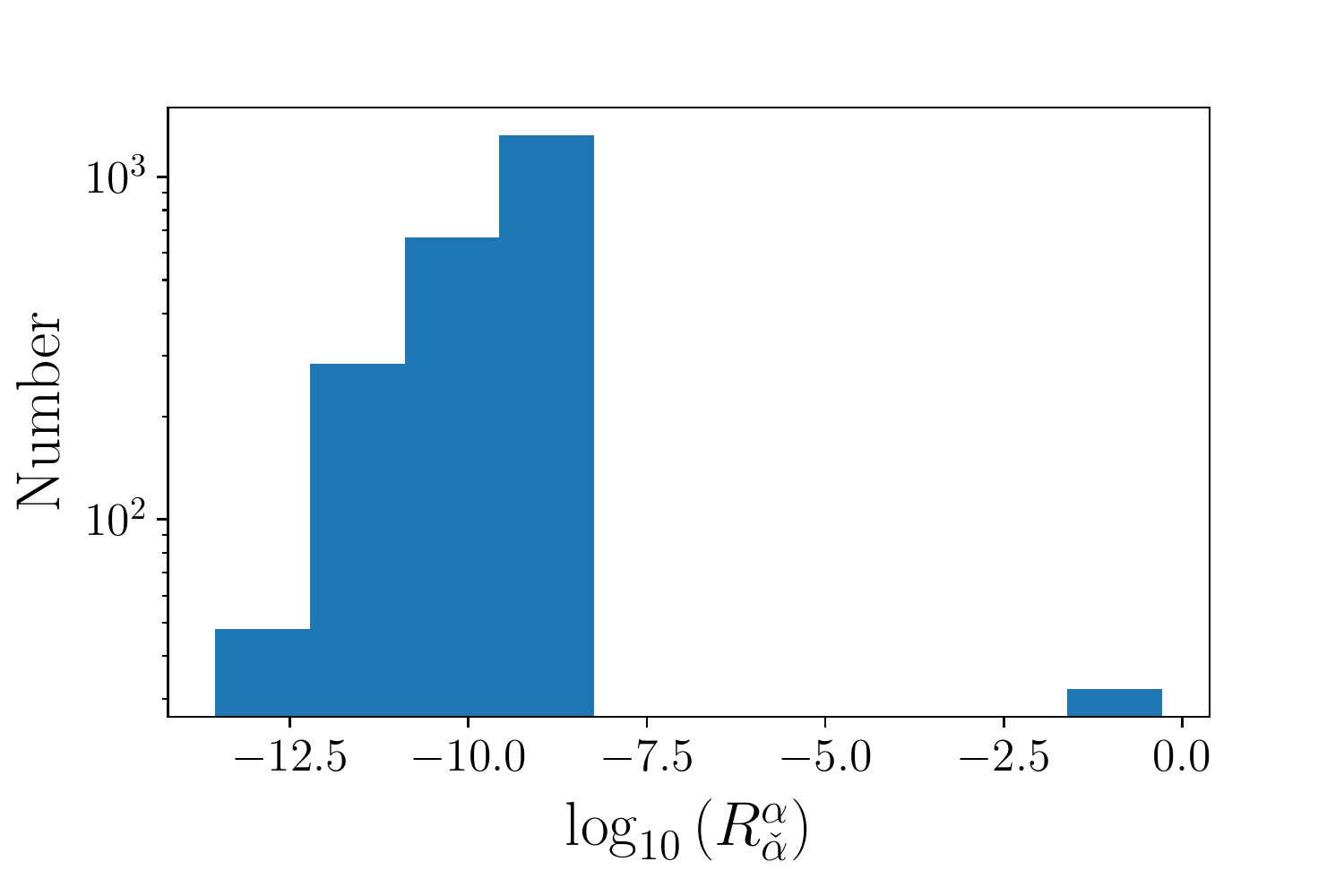}
\caption{
The distribution of $\cR_{\check{\alpha}}^\alpha$ from $B_{\text{max}}$, the geometry with the most flux vacua. This distribution shows $\cR_{\check{\alpha}}^\alpha$ for all choices of $\check{\alpha}$ and $\alpha$ in $B_{\text{max}}$. These ratios are for the most part $ \ll 10^{-8}$, and in fact demonstrate couplings that are even more asymmetric than those found in the Tree ensemble. The peak on the right consists of 32 entries, each due to the $G_2 \times SU(2)$ sectors in $B_{\text{max}}$.
}
\label{fig:bmaxdist}
\end{figure}

While we have mainly considered geometries in the Tree ensemble, we will also briefly discuss the F-theory base geometry which is believed to support the largest number of flux vacua ($\sim 10^{272,000}$), denoted by $B_{\text{max}}$~\cite{Taylor:2015xtz}. This geometry has a non-Higgsable gauge sector of the form $E_8^9 \times F_4^8 \times (G_2 \times SU(2)))^{16}$, and has 98 axions. We again consider gauge-oriented reheatons, and compute $\cR_{\check{\alpha}}^\alpha$. The distribution is shown in Fig.~\ref{fig:bmaxdist}. Clearly this distribution is consistent with the asymmetric couplings that we saw in the Tree ensemble: the ratios $\cR_{\check{\alpha}}^\alpha$ tend to be small. However, from Fig.~\ref{fig:bmaxdist} we see that $B_{\text{max}}$ actually gives rise to couplings that are even more asymmetric than in the Tree ensemble: there are no instances of $\cR_{\check{\alpha}}^\alpha \geq 1$, and the right tail of entries with $\cR_{\check{\alpha}}^\alpha \lesssim 1$ is quite small compared to the bulk. In fact, there are 32 entries in the right tail, which readily yields a simple geometric explanation: the only pairs of gauge groups whose corresponding divisors intersect are the factors of $(G_2 \times SU(2)))$. As there are 16 such factors, there are 32 gauge-oriented reheaton choices whose corresponding divisors intersect a divisor supporting another gauge group. This provides further evidence for our local geometric interpretation.

In conclusion, asymmetric reheating is even easier to arrange in $B_{\text{max}}$, compared to the Tree ensemble. Via this
mechanism, if the reheaton is oriented along an $E_8$ or $F_4$ gauge sector, it will only significantly reheat that sector. If the reheaton is oriented along a $G_2$ or $SU(2)$, then it will reheat that sector as well as an additional $SU(2)$ or $G_2$, respectively.

\subsection{Saxion reheatons}
As a final note in this section we will briefly comment on how our analysis applies to saxion reheatons. From Lagrangian Eq.~\ref{eqn:eff}, we see that the couplings of a given saxion to the gauge fields is actually the same as its axion partner, with $F \wedge F \rightarrow F \wedge \star F$.  In
the context of $\mathcal{N}=1$ string compactifications, since the K\"ahler metric for K\" ahler moduli applies to both their 
saxion and axion components, our analysis thus far applies equally well to the saxion-gauge field couplings. However, since the saxions are not protected by a discrete shift symmetry, we expect the potential of the saxions to be richer than that of the axions after supersymmetry breaking, which could include extra couplings that lead to higher 
decay rates than the dimension five coupling to gauge fields
that we study. Whether asymmetric reheating persists in this
regime for saxions requires a careful analysis, but in the
absence of these additional couplings our result also
holds for saxions. 

\section{Reheated Gauge Groups}
\label{sec:gaugecorr}

\begin{figure}
\includegraphics[width=.5\textwidth]{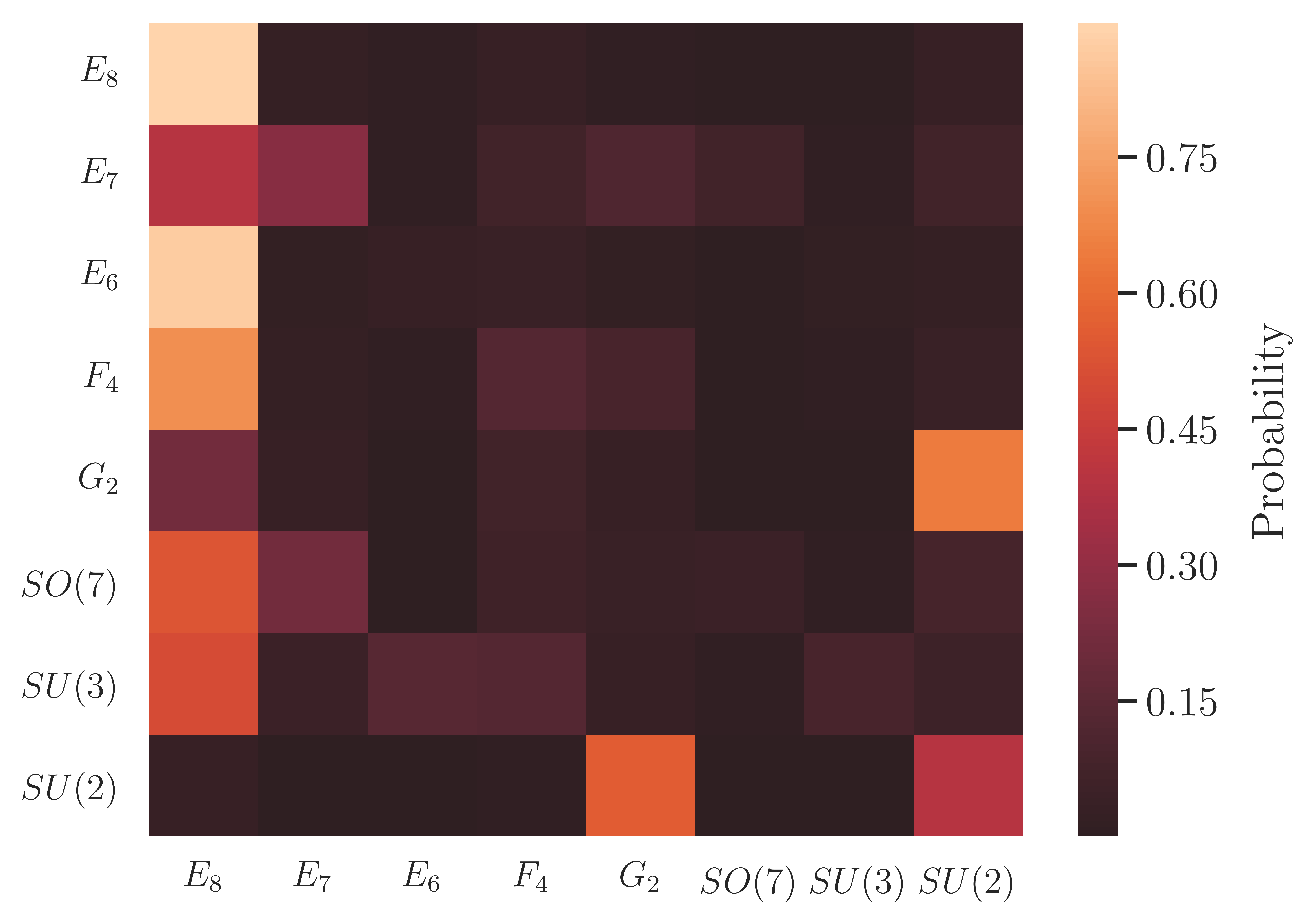}
\caption{Given a $\check{\phi}$ oriented along a particular gauge group, labeled by the vertical axis, this plot shows the additional sectors which have preferred couplings, labeled by the horizontal axis. In this case, asymmetric is defined to be $\cR_{\check{\alpha}}^\alpha \geq 0.1$. While most correlations are with $E_8$, both $G_2$ and $SU(2)$-oriented reheatons also prefers to couple to an additional $SU(2)$ sector. This trend is consistent with the topological distribution from the Tree ensemble. 
}
\label{fig:gauge}
\end{figure}

Thus far, we have demonstrated asymmetric couplings of axions, determined by both topological structure and the position in K\"ahler moduli space. It is natural to ask, what does this imply about correlations between various gauge sectors? In Fig.~\ref{fig:gauge} we show the probability distribution of a $G_{\check{\alpha}}$-oriented reheaton to couple significantly ( $\cR_{\check{\alpha}}^\alpha \geq 0.1$) to a gauge sector $G_\alpha$. The distribution is quite interesting. The axion-reheatons oriented along gauge groups apart from $G_2$ and $SU(2)$ also tend to significantly couple to an $E_8$ universally. However, axion-reheatons oriented along $G_2$ or $SU(2)$ instead tend to significantly couple to another $SU(2)$. As an example let us consider the reheatons oriented along  $SU(2)$, which exists (to high probability) on so-called height-4 rays (see~\cite{Halverson:2017ffz} for further details).  The most significantly couplings are to $G_2$ and other $SU(2)$ gauge group, with a smaller fraction to $E_8$.

From the topological point of view we can consider all of the tree configurations and their corresponding gauge groups, and ask which gauge groups are most likely to be connected to $SU(2)$ via divisor intersections. To do so we make a technical assumption, which is that the initial point over which the tree was placed was defined by a triple intersection of divisors carrying $E_8$ gauge groups (this assumption is of very high probability in the Tree ensemble~\cite{Halverson:2017ffz}). It is then the case that every single $SU(2)$ in the tree ensemble shares an edge with a $G_2$: there is a universal $G_2$ that arises in the first blowup, and $SU(2)$ arises from a blowup along an edge connecting that $G_2$ to a vertex. If another $SU(2)$ is present, whether or not the $\phi^{\check{\alpha}}$ $SU(2)$ intersects with the other $SU(2)$ is a question of the particular tree, and this occurs about $90\%$ of the time. The case of an $E_8$ coupling is similar, and a given $SU(2)$ intersects an $E_8$ at a frequency of about $10\%$. This statistics roughly resemble the plot given in 
Fig.~\ref{fig:gauge}, though of course moduli-dependent data also enter the reheating coefficients, not just topological data. 
With this topological correlation in mind, one should note that the preference of an $E_8$-oriented reheaton to couple significantly to other $E_8$'s should be taken with a grain of salt, as we have not included trees over edges in this analysis. Such trees will generically separate interesting $E_8$ divisor, and we therefore expect that this preference is an artifact of our analysis. 

\begin{figure}
\includegraphics[width=.5\textwidth]{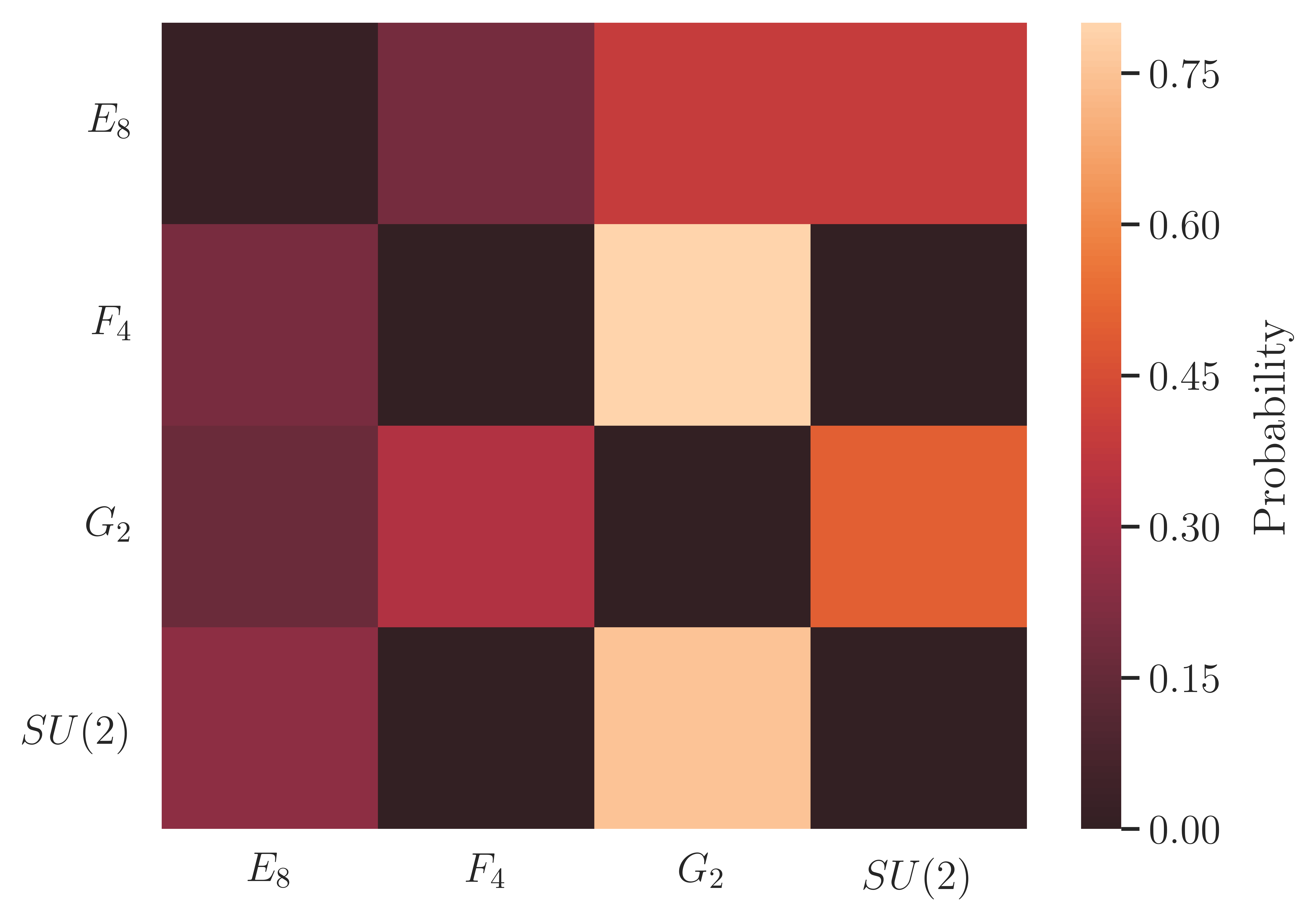}
\caption{The distribution of topological intersections of trees over curves in the Tree ensemble, assuming the curve corresponds to intersecting $E_8$'s. We anticipate the main effect of including curve tree is to shift the preference of an $SU(2)$-oriented reheaton from another $SU(2)$ to $G_2$, and to virtually eliminate the preference of an $E_8$-oriented reheaton to couple to another $E_8$. 
}
\label{fig:gaugecurve}
\end{figure}

As a final consideration, we note that we have concentrated on trees over points, but one could also consider trees over curves as well. The structure is not nearly as rich, as there are only 82 allowed trees over curves in the Tree ensemble, but we expect this to alter the distribution shown in Fig.~\ref{fig:gauge} somewhat. Above we gave evidence that the distribution for the $SU(2)$-aligned reheaton roughly coincides with the distribution of topological intersections of gauge group divisor in the Tree ensemble. With that in mind, we will estimate the effect that trees over edges will have on our correlations by considering such intersections for trees over curves. We make the high probability assumption that the two divisors, whose intersection defines the curve over which the tree is placed, both carry an $E_8$ gauge group. Using this high-probability technical assumption the probability distribution is shown in Fig~\ref{fig:gaugecurve}. This assumption restricts the type of gauge groups carried by the divisors in the tree to those that appear in Fig~\ref{fig:gaugecurve}. From this distribution we anticipate the main effect of including curve tree is to shift the preference of an $SU(2)$-oriented reheaton from another $SU(2)$ to $G_2$, and to virtually eliminate the preference of an $E_8$-oriented reheaton to couple to another $E_8$.

\section{Discussion} 
\label{sec:discuss}

In this work we demonstrated that asymmetric
axion 
reheating arises in F-theory compactifications
that have large numbers of axions and gauge
sectors.
We computed
couplings relevant for reheating, in particular the couplings
of axions to gauge groups in a large ensemble of F-theory geometries and also the F-theory
geometry with the most flux vacua,
both of which generically exhibit large
numbers of axions and gauge sectors. 

 The distributions of coupling ratios are  non-trivial. One main result is that
gauge-oriented axions couple to $O(1\%)$
of the total gauge sectors with $O(1)$
couplings, while randomly drawn combinations
of those axions couple to $O(10\%)$.
In the gauge-oriented case, the peak of
the distribution is far below zero.
Asymmetric reheating occurs in both cases,
and leads to reheating that is
more asymmetric than in bottom-up models
obtained from either $O(1)$ Wilson 
coefficients or random EFTs.

Another result in the gauge-oriented axion case
 is that the leading
couplings admit an interpretation that
is local in the string geometry. Specifically,
their ratios of reheating couplings 
depend critically on local intersection structure.  These ratios are a function of both discrete topological structure and continuous K\"ahler moduli. 

Of course, this is only a first step in understanding reheating dynamics in broad classes of string compactifications. For instance, we have not included additional couplings of the axions to the saxions (and other moduli), such as those that would be generated by a non-perturbative superpotential. Such couplings could in principle be dangerous for asymmetric reheating, as one could imagine a scenario in which saxions become excited
and themselves contribute to reheating.\footnote{We thank Liam McAllister for discussion on this point.} 

In addition, we have not yet identified candidate reheatons in this setup, but have instead taken an agnostic
approach. One option is that the reheaton is not connected to inflation, but is instead
simply a scalar field that comes to dominate
the energy density of the universe and
decays prior to nucleosynthesis; this
is relatively simple, but depends on
moduli stabilization and supersymmetry breaking.
Another possibility is that reheaton is the
inflaton, which requires identifying which
axion direction begins to oscillate
at the end of inflation. This hard problem
requires control over inflationary
trajectories in high-dimensional
scalar potentials.  Both
cases require control over the scalar potential and need to be computed with care. 

\vspace{.5cm}
Finally, we return to one of our motivations:
reheating too many degrees of freedom
in string compactifications can lead to
severe cosmological problems that
disagree with observations, such
as the overproduction of dark matter. Though
avoiding these problems is model-dependent,
our result that asymmetric reheating
arises naturally in large string ensembles
could play a major role in obtaining
realistic post-inflationary cosmologies.

\vspace{.2cm}
\noindent{\bf Acknowledgements.} We thank JiJi Fan, Liam McAllister, Fernando Quevedo, Fabian Ruehle, Jessie Shelton, and Scott Watson for useful discussion. We acknowledge generous support provided by the Northeastern University Discovery Cluster. J.H. is supported by NSF grant PHY-1620526. B.N. is supported by NSF grant PHY-PHY-1620575.

\bibliography{refs}

\end{document}